\documentclass[12pt]{article}
\usepackage{epsf}
\topmargin
 -.5cm
\def\timesbox{\hbox{$\scriptscriptstyle\times$}}
\def\ant{ {{\lower 1ex  \timesbox} \atop {\raise 1.5ex
\timesbox}}}
\newcommand{\Zop}{{\hbox{ Z\kern-1.6mm Z}}}
\newcommand{\beq}{\begin{equation}}
\newcommand{\eeq}{\end{equation}}
\newcommand{\bea}{\begin{eqnarray}}
\newcommand{\eea}{\end{eqnarray}}
\newcommand{\ra}{\rangle}
\newcommand{\la}{\langle}
\newcommand{\lt}{\left}
\newcommand{\rt}{\right}
\newcommand{\Iop}{\relax{\rm I\kern-.18em I}}
\newcommand{\one}{{\hbox{ 1\kern-1.2mm l}}}
\newcommand{\T}{{\cal T}}
\newcommand{\Tt}{{\tilde {\cal T}}}

\newcommand{\dt}{\delta}

\newcommand{\del}{\partial}

\newcommand{\A}{{\cal A}}

\newcommand{\CH}{{\cal H}}

\newcommand{\D}{\Delta}

\newcommand{\eps}{\epsilon}

\newcommand{\s}{\sigma}

\newcommand{\sectiono}[1]{\section{#1}\setcounter{equation}{0}}
\newcommand{\subsectiono}[1]{\subsection{#1}}

\begin{document}

{}~
{}~
\hfill\vbox{\hbox{IMSc/2009/02/03}}
\break

\vskip 2cm

\centerline{\Large \bf Superstrings in type IIB R-R plane-wave in}
\centerline{\Large \bf semi-light-cone gauge and conformal invariance}

\medskip

\vspace*{4.0ex}

\centerline{\large \rm Partha Mukhopadhyay }

\vspace*{4.0ex}

\centerline{\large \it Institute of Mathematical Sciences}
\centerline{\large \it C.I.T. Campus, Taramani}
\centerline{\large \it Chennai 600113, India}

\medskip

\centerline{E-mail: parthamu@imsc.res.in}

\vspace*{5.0ex}

\centerline{\bf Abstract}
\bigskip

We reconsider the analysis done by Kazama and Yokoi in arXiv:0801.1561 (hep-th). We find that although the right vacuum of the theory is the one associated to massless normal ordering (MNO), phase space normal ordering (PNO) plays crucial role in the analysis in the following way. While defining the quantum energy-momentum (EM) tensor one needs to take into account the field redefinition relating the space-time field and the corresponding world-sheet coupling. We argue that for a simple off-shell ansatz for the background this field redefinition can be taken to be identity if the interaction term is ordered according to PNO. This definition reproduces the correct physical spectrum when the background is on-shell. We further show that the right way to extract the effective equation of motion from the Virasoro anomaly is to first order the anomaly terms according to PNO at a finite regularization parameter $\eps$ and then take the $\eps \to 0$ limit. This prescription fixes an ambiguity in taking the limit for certain bosonic and fermionic contributions to the Virasoro anomaly and is the natural one to consider given the above definition of the EM tensor.

\newpage

\tableofcontents

\baselineskip=18pt

\section{Introduction}
\label{s:intro}

Green-Schwarz superstring is usually quantized in light-cone gauge
\cite{gsw}, in which case the world-sheet theory does not have conformal invariance. This invariance can however be preserved in
semi-light-cone gauge \cite{slc, slc-anomaly, berkovits04}, where the $\kappa$-symmetry is fixed along with the conformal gauge for the bosonic part. A BRST method for this gauge was discussed by Berkovits and Marchioro in \cite{berkovits04} for flat background. More recently the same approach was considered in \cite{kazama} by Kazama and Yokoi (KY) to study type IIB  string theory in R-R plane-wave background\cite{blau, metsaev01, metsaev02}. The main motivation was to establish conformal invariance of the theory\footnote{See \cite{CFT-RR} for other CFT  approaches for this background. More general R-R pp-waves have been discussed in \cite{RR-pp}.}. Here we reconsider their analysis to understand certain puzzling issues that will be explained as we go on.

Unlike in light-cone gauge \cite{metsaev01, metsaev02}, in this case the world-sheet theory is interacting which makes the standard canonical quantization difficult \cite{kazama} (see also \cite{chizaki}). However, a phase-space (operator) method was used by KY to compute the Virasoro algebra. Since such a computation involves calculating only equal-time commutators, it can in principle be done without solving for time-evolution once all the operators are expressed in terms of the phase-space modes\footnote{The equal-time commutators/anti-commutators of the full set of phase-space modes leads to an algebraic structure that can be used to define a Hilbert space and is same in both the flat background and the R-R plane-wave.}. However, one needs to choose a normal ordering prescription. The prescription that
gives the right theory for flat background has been called massless
normal ordering (MNO). In \cite{kazama} the following result was shown for R-R plane-wave:
\bea
&& \hbox{\it If the theory is defined with MNO, then the Virasoro
  algebra is not satisfied due}\cr
&& \hbox{\it to non-zero anomalous terms. However, this anomaly
  vanishes if the theory is}\cr
&& \hbox{\it defined with a different prescription called {\it
    phase-space normal ordering} (PNO).} \cr &&
\label{KYresult}
\eea
It was therefore concluded that PNO is the right choice for R-R
plane-wave.

In a previous work \cite{mukhopadhyay08} we pointed out that such a result is very surprising from the point of view of the universality property of pp-waves as discussed in \cite{mukhopadhyay06}. Because of this property one expects that the vacuum of the world-sheet theory in a pp-wave background should correspond to the same normal ordering prescription as in flat background. To understand this in a simpler setting we considered the simplest class of off-shell pp-waves in bosonic string theory where only the lower $++$ component of the metric is switched on. By defining the theory with MNO it was shown that the Virasoro algebra is satisfied and the correct physical spectrum is reproduced when the background is on-shell.

In this work we address the similar issues in the original context of R-R plane-wave. To this end we consider the same off-shell metric as above along with a constant five-form R-R flux. Because of the flux the effective equation of motion for the background becomes slightly more complicated. This makes the results of \cite{mukhopadhyay08} partially invalid for the present case in the following sense. Although it is still true that the right vacuum of the theory is the one associated to MNO as expected, definition of the quantum energy-momentum (EM) tensor needs to be modified. Such a modification does not affect the analysis of \cite{mukhopadhyay08},
but it does in the present case where a flux is turned on. Moreover,
extracting the effective equation of motion\footnote{See
  \cite{lovelace, fradkin, callan, sen} for a partial list of original
  references on this subject.} from the Virasoro anomaly becomes more subtle. Below we will discuss these two issues in further detail.

We discuss the definition of the quantum EM tensor in section
\ref{s:EMtensor}. According to this definition the {\it free part} is ordered according to MNO, but the {\it interaction part} according to PNO. The reason for this definition can be understood as follows. The fields appearing in the space-time effective theory and the
corresponding world-sheet couplings are in general related by field
redefinition. We argue that for the restricted off-shell ansatz considered here this field redefinition can be taken to be identity if the interaction term is ordered according to PNO. One is free to order this term according to MNO as well, but in that case a non-trivial field redefinition needs to be considered. We show in subsection \ref{ss:spectrum} that with this definition the correct physical spectrum is reproduced for the on-shell R-R plane-wave.

We discuss the computation of Virasoro anomaly in section
\ref{s:Vir}. There are three anomaly terms denoted by
$\A^R(\s,\s')$, $\A^L(\s,\s')$ and $\A(\s,\s')$ resulting from the
equal-time commutators $[\T(\s),\T(\s')]$, $[\Tt(\s), \Tt(\s')]$ and
$[\T(\s), \Tt(\s')]$ respectively. Here $\s$ is the world-sheet space coordinate\footnote{Throughout the paper we will suppress
  the world-sheet time coordinate $\tau$ in the arguments.
All our analysis are understood to be done at the same $\tau$.} and
$\T(\s)$ and $\Tt(\s)$ are the right and left moving EM tensor
components respectively. In the present method of computation each of the above anomaly terms receives contribution from an infinite number of phase-space modes and the actual computation is done in a
regularized theory with a finite cutoff $\eps$. We show that certain
bosonic and fermionic contributions to the anomaly terms develop an
ambiguity when we take the $\eps \to 0$ limit. This ambiguity is
related to whether we order the operators according to MNO or
PNO at a finite cutoff. It turns out that the correct equation of
motion emerges in a satisfactory manner if we order the terms according to PNO before taking the limit. This is also required by the definition of EM tensor discussed above.

It is important to note that consideration of PNO as discussed
above is not related to the choice of vacuum. Although the interaction term in the EM tensor is ordered according to PNO, the correct spectrum is reproduced when we define the vacuum to be the one associated to MNO. In fact the spectrum contains negative dimensions, as expected \cite{mukhopadhyay08}, if we choose the PNO-vacuum instead (see appendix \ref{a:PNO}). In the context of Virasoro algebra let us consider the fermionic contribution to the Virasoro anomaly. It can be shown using an alternative argument, which does not require us to order the terms in any particular way, that the result is same as obtained in PNO-prescription even in the MNO-vacuum.

We conclude in section \ref{s:conclusion} and keep some technical
details in various appendices. Some of the technical results described here were already known in \cite{kazama}. However, our analysis and interpretations are very different.

\section{Energy-momentum tensor}
\label{s:EMtensor}

\subsection{Definition and field redefinition}
\label{s:def}
The relevant details of the semi-light-cone quantization and the regularization procedure have been described in appendix \ref{a:quant} and \ref{a:reg} respectively from where we will borrow various notations for the discussion to follow. We define the EM tensor in the usual manner:
$T_{ab} = - {4\pi \over \sqrt{-g}} {\delta S \over \delta g^{ab}}$. The right and left moving components are given by: $\T ={1\over 2} (T_{00} - T_{01})$ and $\tilde \T ={1\over 2} (T_{00} + T_{01})$ respectively. The quantum operators are given by the following expressions:
\bea
\T &=& \T^{(0)} + \dt \T~, \quad \tilde \T = \tilde \T^{(0)} +
\dt \T~,
\label{calT}
\eea
where the {\it free parts} relevant to the flat background are given by\cite{berkovits04} (see also \cite{slc-anomaly})\footnote{We will use the
  notations: $A.B=\eta_{\mu \nu} A^{\mu} B^{\nu}$ and $\vec A. \vec B
  = A^I B^I$.},
\bea
\T^{(0)} &=& : \lt({1\over 2} \Pi. \Pi - {i\over 2} ( S\del_{\s}S)+
\xi \del_{\s}^2 \ln \Pi^{+}\rt): +1 ~, \cr
\tilde \T^{(0)} &=& :\lt( {1\over 2} \tilde \Pi. \tilde \Pi  + {i\over
  2} (\tilde S \del_{\s} \tilde S) + \xi \del_{\s}^2 \ln \tilde
\Pi^{+} \rt): + 1~,
\label{calT0}
\eea
where $:~:$ refers to MNO. The last terms inside the round brackets arise from non-covariant gauge fixing. The normal ordering constant $1$ and $\xi = -{1\over 2}$ have been fixed in appendix \ref{a:constant} from the condition that $\T^{(0)}$ and $\Tt^{(0)}$ satisfy the standard form of the Virasoro algebra with central charge $26$. This cancels with $-26$ coming from the $(b,c)$ ghost system which turns out to be the only propagating ghost degrees of freedom. The {\it interaction part} is given by: $\dt \T = \dt \T_B+\dt \T_F$
where,
\bea
\dt \T_B &=& - {1\over 2} \chi^2 K(\vec X)~, \quad
\dt \T_F = {i\mu \over 2 \sqrt{\pi T}} ~\chi (S\Sigma \tilde S) ~,
\label{DeltaT}
\eea
where $\chi(\s)= \sqrt{\Pi^+ \tilde \Pi^+(\s)}$, $\Sigma = \s^{1234}$
and,
\bea
K(\vec x) = -\mu^2 \vec x^2~.
\label{quad}
\eea
As mentioned earlier the interaction part is ordered according to PNO. For both the bosonic and fermionic interactions in (\ref{DeltaT}) PNO gives rise to operators without any non-trivial re-ordering of the modes. Given the anti-commutators below eqs.(\ref{mode-exp}), this is obvious for $\dt \T_F$. For $\dt \T_B$ see the discussion below eqs.(\ref{D-result}).

The above definition can be understood to relate, in a particular way, the field in the space-time effective theory and the corresponding world-sheet coupling. To see that more explicitly let us consider an off-shell ansatz for the background where the R-R flux is still constant, but the metric is given by,
\bea
ds^2 = 2 dx^+ dx^- + K(\vec x) (dx^+)^2 + d\vec x. d \vec x~,
\label{off-shell}
\eea
where $K(\vec x)$ is arbitrary. This background is characterized by
the Fourier transform $\tilde K(\vec p) = \int {d\vec x \over
  (2\pi)^{d/2} } K(\vec x) e^{-i\vec p.\vec x}$. According to our definition the same Fourier transform is used as the world-sheet coupling in $\dt \T_B$ in (\ref{DeltaT}), i.e.
\bea
K(\vec X(\s)) = \int {d\vec p \over (2\pi)^{d/2} }~ \tilde K(\vec p)
e^{i\vec p.\vec X}(\s)~,
\label{K-PNO}
\eea
provided the exponential operator is unordered (i.e. ordered according to PNO). We could alternatively define the above operator according to MNO,
\bea
K(\vec X(\s)) = \int {d\vec p \over (2\pi)^{d/2}}  ~\tilde K'(\vec p)
: e^{i\vec p. \vec X}(\s) : ~.
\label{K-MNO}
\eea
But in that case the Fourier transform $\tilde K'(\vec p)$ has to be
identified with the one characterizing the background
(\ref{off-shell}) up to the following field redefinition,
\bea
\tilde K'(\vec p) = e^{-{\vec p^2\over 2} D_{\eps}(0)} \tilde K(\vec
p)~,
\label{field-redef}
\eea
where $\eps$ is the regularization parameter and $\eta^{\mu \nu}
D_{\eps}(\Delta)$ is the bosonic ``equal time propagator'' as
discussed in appendix \ref{a:reg}. This relation is simply obtained by equating the operators in (\ref{K-PNO}) and (\ref{K-MNO}) and using $e^{i\vec p. \vec X}(\s)= e^{-{\vec p^2 \over 2} D_{\eps}(0)}
:e^{i\vec p. \vec X}(\s):$.

\subsection{Physical spectrum}
\label{ss:spectrum}

Given the above definition of the quantum EM tensor we will now
compute the physical spectrum for the quadratic profile in
(\ref{quad}) and show that it reproduces the right answer. We will
follow the method of \cite{mukhopadhyay08} which requires us to have a CFT formulation. Here we will simply assume
the conformal invariance which will be established in the next
section.

Unless stated otherwise we will mostly follow the notations of
\cite{mukhopadhyay08}. We take the quadratic space-time action to be
as given in eq.[2.25]\footnote{Equation numbers kept in square
  brackets will refer to equations from \cite{mukhopadhyay08}.} where the inner product is understood to be the hermitian inner product. The orthonormal basis
states spanning the transverse Hilbert space $\CH_{\perp}$ are given
by $|\{N\}, p, \eta) = c_1 \tilde c_1| \{N\}, p, \eta\ra$ such that
$|\{N\}, p, \eta)$ satisfies the conditions [2.21]. Here $\{N\}$
represents a set of four
sets of integers, namely: $\{\{N^B_{In}\}, \{\tilde N^B_{In}\},
\{N^F_{an}\}, \{\tilde N^F_{an}\} \}$ such that $|\{N\}, p, \eta \ra$
is proportional to:
\bea
\prod_{n>0, I, a} (\Pi^I_{-n})^{N^B_{In}}
(\tilde \Pi^I_{-n})^{\tilde N^B_{In}}  (S^a_{-n})^{N^F_{an}} (\tilde
S^a_{-n})^{\tilde N^F_{an}} |p, \eta \ra~,
\label{basis1}
\eea
where the momentum $p^{\mu}$ of the ground states $|p,\eta \ra =
e^{ip.X_0}|\eta \ra$ is normalized as:
$\Pi^{\mu}_0 |p, \eta \ra =  {p^{\mu} \over 2\sqrt{\pi T}} |p, \eta
\ra$. From the linearized equation of motion one obtains the following eigenvalue equation:
\bea
{\cal S}_2 |\Psi_{\perp}\ra = 0~,
\eea
where,
\bea
{\cal S}_2 &=& L_0 +\tilde L_0 -2 ~, \quad \quad |\Psi_{\perp}\ra =
\sum'_{\eta, \{N\}} \int dp ~\tilde \psi_{\eta, \{N\}}(p) |\{N\}, p,
\eta\ra ~.
\label{calS2}
\eea
The prime on the summation refers to the condition $(L_0-\tilde
L_0)=0$. The normal ordering constant in $L_0$ and $\tilde L_0$ has
been fixed in appendix \ref{a:constant} which precisely cancels the
ghost contribution of $-2$ in the first equation of
(\ref{calS2}). Using this we get the following result inside
$\CH_{\perp}$:
\bea
{\cal S}_2 &=& {\cal S}^0_2 + {\cal S}_2^{\neq 0}~, \cr
{\cal S}_2^0 &=&  {\alpha' \over 2} \lt[ p^2 + (\mu p^+)^2 X^I_0 X^I_0
+ 2i \mu p^+ (S_0\Sigma \tilde S_0) \rt]~, \cr
{\cal S}_2^{\neq 0} &=& \sum_{n>0} \lt[ \Pi^I_{-n} \Pi^I_n + \tilde
\Pi^I_{-n} \tilde \Pi^I_n + n (S_{-n}S_n) + n (\tilde S_{-n} \tilde
S_n) \rt] \cr
&&+ \sum_{n\neq 0} \lt[ {m^2\over 4n^2}  (\Pi^I_{-n} \Pi^I_n + \tilde
\Pi^I_{-n} \tilde \Pi^I_n) -{m^2\over 2n^2} \Pi^I_n \tilde \Pi^I_n +im
(S_{n}\Sigma \tilde S_{n})\rt] ~,
\label{calS2perp}
\eea
where $m=\alpha' \mu p^+$. To diagonalize the bosonic part of ${\cal
  S}_2^{\neq 0}$ we first define the operators $\alpha^I_n$ and
$\tilde \alpha^I_n$ ($n\neq 0$) precisely in the same way as in
\cite{mukhopadhyay08} by replacing $m_I \to m$ in eqs.[2.30]. For the fermionic part one defines $U_n$ and $\tilde U_n$ ($n\neq 0$) in the
following way:
\bea
S_n = c_n \lt(U_n - i{w_n-n \over m} \Sigma \tilde U_{-n} \rt)~, \quad
\tilde S_n = c_n \lt(\tilde U_n + i{w_n-n \over m} \Sigma U_{-n} \rt)~,
\label{U-S}
\eea
where $w_n = n \sqrt{1+{m^2\over n^2}}$ and $c_n = {1\over
  \sqrt{1+\lt(w_n-n \over m \rt)^2}}$. With this definition one can
establish: $\{U^a_m, U^b_n\} = \{\tilde U^a_m,\tilde
U^b_n\}=\delta^{ab} \delta_{m+n,0}$, $\{U^a_m,\tilde U^b_n\}=0$. Next we define new vacua $|p, \eta\ra\ra$ such that
$(\alpha^I_n,  \tilde \alpha^I_n, U^a_n, \tilde U^a_n) |p, \eta \ra
\ra = 0 ~, \quad \forall n>0$, and a new basis $|\{N\},p,\eta \ra \ra$
which can be obtained by replacing $\Pi$ and $S$ oscillators in the
expression (\ref{basis1}) by the corresponding $\alpha$ and $U$ oscillators respectively. Expanding $|\Psi_{\perp}\ra$ in this new basis one gets the following expected result\cite{metsaev02}:
\bea
{\cal S}_2^{\neq 0} &=& \sum_{n>0} \lt[ \alpha^I_{-n} \alpha^I_n +
\tilde \alpha^I_{-n} \tilde \alpha^I_n + w_n \lt(U^a_{-n}U^a_n+\tilde
U^a_{-n} \tilde U^a_n \rt)\rt]~.
\label{calS2diag}
\eea
Such a change of basis produces c-number contributions of $d
\sum_{n>0} (w_n-n)$ each for the bosonic and the fermionic parts but
with opposite signs so that they cancel each other.

We mainly followed \cite{mukhopadhyay08} for the above derivation. However, notice that unlike the definition considered here, $\dt \T_B$ was ordered according to MNO in that work. Had we done the same thing here, the bosonic part of the last line in (\ref{calS2perp}) would have been replaced by,
\bea
&&\sum_{n>0, I} {m^2 \over 2 n^2} \lt(\Pi^I_{-n}\Pi^I_n + \tilde
\Pi^I_{-n} \tilde \Pi^I_n - \Pi^I_n \tilde \Pi^I_n - \Pi^I_{-n} \tilde
\Pi^I_{-n} \rt)~, \cr
&=& \lt[ \sum_{n \neq 0, I} {m^2 \over 4 n^2} \lt(\Pi^I_{-n}\Pi^I_n +
\tilde \Pi^I_{-n} \tilde \Pi^I_n \rt) - {m^2 \over 2 n^2}\Pi^I_n
\tilde \Pi^I_n  \rt] +  Z(m^2)~,
\eea
where the term in the square bracket is the operator ordered with PNO as considered here. But there is an additional contribution,
\bea
Z(m^2) &=& - {dm^2 \over 2}  \sum_{n>0} {1\over n}~,
\label{DeltaL0}
\eea
which can easily be recondensed as\footnote{We will use the notation
  $\oint $ for the definite integral $\int_0^{2\pi}{d\s\over 2\pi}$.}
$d\mu^2 D_{\eps}(0)\oint \chi(\s)^2$ evaluated inside ${\cal
  H}_{\perp}$. This additional negative contribution (leading to a
wrong spectrum) is coming due to the fact that we have wrongly
identified $\tilde K'(\vec p)$ in eq.(\ref{K-MNO}) (instead of $\tilde K(\vec p)$ in eq.(\ref{K-PNO})) with the Fourier transform $(2\pi)^{d/2} \mu^2 \del_{\vec p}.\del_{\vec p} \delta(\vec p)$ which corresponds to the quadratic profile (\ref{quad}). This is precisely the reason why the (unwanted) factor $Z(m_I^2) = -{1\over 2}\lt(\sum_Im_I^2\rt)\sum_{n>0}{1\over n}$ appeared in the zero-point energy in eq.[2.32]. In that case this factor is zero as $\sum_Im_I^2 =0$ because of the simple on-shell condition in [2.3] (see also [2.14], [2.15]). However, in a more complicated background, such as the one considered in this paper where a flux is turned on, such a contribution is not zero.

\section{Virasoro anomaly}
\label{s:Vir}

\subsection{Equation of motion}
\label{ss:eom}

In this section we will discuss the computation of Virasoro
algebra. In particular, we will ask how to extract, from the Virasoro anomalies, the supergravity equation of motion which, for the present off-shell ansatz, is simply given by,
\bea
-{1\over 2} \vec \del^2 K = d\mu^2~.
\label{eom}
\eea
It is only when the above equation is satisfied we would expect all
the three equal time commutators $[\T(\s),\T(\s')]$, $[\Tt(\s),
\Tt(\s')]$ and $[\T(\s), \Tt(\s')]$ to satisfy the standard
expressions for the Virasoro algebra (see for example
\cite{mukhopadhyay08})\footnote{The definitions of $\T$($\T^{(0)}$)
  and $\Tt$($\Tt^{(0)}$) considered here are interchanged with respect
  to that in \cite{mukhopadhyay08}.}. However for a generic off-shell configuration each one is supposed to contain an anomaly term. Following \cite{mukhopadhyay08} we use the Virasoro algebra
satisfied by $\T^{(0)}(\s)$ and $\tilde \T^{(0)}(\s)$ to derive the
following expressions for these anomaly terms:
\bea
{\cal A}^R(\s,\s') &=& A^R(\s,\s') + [\dt \T_F(\s), \dt
\T_F(\s')]~,  \cr
{\cal A}^L(\s,\s') &=& A^L(\s,\s') + [\dt \T_F(\s), \dt
\T_F(\s')]~, \cr
{\cal A}(\s,\s') &=& A(\s,\s') + [\dt \T_F(\s), \dt \T_F(\s')]~,
\label{anomalies}
\eea
where we have used: $[\dt \T_B(\s), \dt \T_B(\s')] =0$ (see
\cite{mukhopadhyay08}) and,
\bea
A^R(\s,\s') &=& [\T^{(0)}(\s), \dt \T(\s')] + [\dt \T(\s),
\T^{(0)}(\s')] + 4 \pi i\dt \T (\s) \delta'(\D) + 2\pi i \del_{\s}
\dt \T(\s) \delta (\D) ~, \cr
A^L(\s,\s') &=& [\tilde \T^{(0)}(\s), \dt \T(\s')] + [\dt
\T(\s), \tilde \T^{(0)}(\s')]  - 4 \pi i\dt \T (\s) \delta'(\D) -
2\pi i \del_{\s} \dt \T(\s) \delta (\D)~, \cr
A(\s,\s') &=& [\tilde \T^{(0)}(\s), \dt \T(\s')] + [\dt \T(\s),
\tilde \T^{(0)}(\s')]~,
\label{bosonic-anomalies}
\eea
where $\D = \s-\s'$.
We will now show that the condition that all the anomaly terms in
(\ref{anomalies}) be zero is same as (\ref{eom}). We borrow the
following results derived in subsection \ref{ss:ambiguity}:
\bea
A^R(\s,\s') = A^L(\s,\s')=A(\s,\s')= {i\over 4T} \lt(2 \vec \del^2
\dt \T_B(\s)
\delta'(\Delta) + \del_{\s} \vec \del^2 \dt \T_B(\s)
\delta(\Delta) \rt)~,
\label{A-results}
\eea
and
\bea
[\dt \T_F(\s), \dt \T_F(\s')]= -{i\over 4 T}
\lt(2 d \mu^2 \chi^2 (\s) \delta'(\D) + d\mu^2 \del_{\s} \chi^2
(\s) \delta (\D) \rt)~,
\label{expected}
\eea
such that,
\bea
{\cal A}^R(\s,\s') ={\cal A}^L(\s,\s')= {\cal A}(\s,\s') = {i\over
  4T}\lt[ 2 E(\s) \delta'(\D) + \del_{\s} E(\s)
\delta(\D)\rt]~,
\label{anomaly-results}
\eea
where,
\bea
E(\s) = \vec \del^2 \dt \T_B(\s) - d\mu^2 \chi^2(\s) ~.
\label{Es}
\eea
Using $\vec \del^2 \dt \T_B = -{1\over 2} \chi^2 \vec
\del^2 K$ we conclude that vanishing of the anomaly terms implies,
\bea
-{1\over 2} \vec \del^2 K(\s) = d\mu^2~,\quad {\rm i.e.} \quad {1\over
  2} \vec k^2 \tilde K(\vec k) = (2\pi)^{d/2} d \mu^2 \delta(\vec k)~,
\label{operator-eom}
\eea
where the last equation is the Fourier transform of (\ref{eom}).

\subsectiono{Ambiguity in computing Virasoro anomaly and PNO -
  prescription}
\label{ss:ambiguity}

The operator $E(\s)$ in eq.(\ref{Es}) receives contribution from two
sources - the first term i.e. the bosonic contribution comes from the anomaly terms in eqs(\ref{A-results}) and the second term comes from
the commutator $[\dt \T_F(\s), \dt \T_F(\s')]$ in the fermionic
sector. As mentioned in appendix \ref{a:reg}, we do the actual
calculations in the regularized theory with a finite cutoff  $\eps$ and take the $\eps \to 0$ limit at the end. Here we will show that following this procedure one encounters ambiguities in computing certain contributions both in the bosonic and the fermionic sectors. This ambiguity originates from the fact that ordering operators according to MNO and PNO in the cutoff theory and then taking the $\eps \to 0$ limit produces different answers. It turns out that the latter, which is the natural one to follow given our definition of the EM tensor, gives rise to the right space-time equation of motion. Below we will consider the two sectors separately.

\vspace{.3in}
\noindent
{\bf \large Bosonic sector}

\noindent
We have shown in appendix \ref{a:proof} that given the expressions in (\ref{bosonic-anomalies}), the bosonic anomaly terms take the following forms:
\bea
A^R(\s,\s') &=& C^B(\s,\s') - C^B(\s',\s) ~, \quad
A^L(\s,\s') = \tilde C^B(\s,\s') - \tilde C^B(\s',\s) ~, \cr
A(\s,\s') &=& C^B(\s,\s') - \tilde C^B(\s',\s) + \pi i \chi^2(\s) \del_{\s}K(\vec X(\s))
\delta(\D)~,
\label{A-forms}
\eea
where,
\bea
C^B(\s,\s') &=& \lt[t(\s), \dt \T_B(\s') \rt]~, \quad \tilde C^B(\s,\s')
= \lt[\tilde t(\s), \dt \T_B(\s') \rt]~,
\eea
where $t(\s)$ and $\tilde t(\s)$ are defined in
eqs.(\ref{calT0-parts}). Therefore the relevant commutators that we
need to compute at finite $\eps$ are given by:
\bea
[:\Pi^I\Pi^I(\s):, e^{i\vec k.\vec X}(\s')]_{\eps} &=&
\sqrt{\pi \over T} \delta_{\eps}(\D)
\lt(\Pi_{\eps}^I(\s)e_{\eps}^{i\vec k.\vec X}(\s')
+ e_{\eps}^{i\vec k.\vec X}(\s')\Pi_{\eps}^I(\s) \rt)~, \cr
[:\tilde \Pi^I\tilde \Pi^I(\s):, e^{i\vec k.\vec X}(\s')]_{\eps} &=&
\sqrt{\pi \over T} \delta_{\eps}(\D)
\lt(\tilde \Pi_{\eps}^I(\s)e_{\eps}^{i\vec k.\vec X}(\s') +
e_{\eps}^{i\vec k.\vec X}(\s') \tilde \Pi_{\eps}^I(\s) \rt)~.
\label{Pi2-e-comm}
\eea
Notice that $A^R(\s,\s')$ and $A^L(\s,\s')$ in (\ref{A-forms}) have
anti-symmetrization in $\s$ and $\s'$. Therefore because of the
$\delta_{\eps}(\D)$ factor in eqs.(\ref{Pi2-e-comm}), classically
these anomalies are zero in the $\eps \to 0$ limit as expected. However, in the quantum theory we are supposed to first order the right hand sides of (\ref{Pi2-e-comm}) according to some normal ordering prescription before taking the limit. We will see that
the final result is different if we order the operators according to
MNO and PNO.

Let us first consider the right moving sector. In our convention the
two normal ordering prescriptions act in the same way in this sector
and therefore there is no ambiguity. The result can be written as:
\bea
\Pi_{\eps}^I(\s)e_{\eps}^{i\vec k.\vec X}(\s') +
e_{\eps}^{i\vec k.\vec X}(\s')\Pi_{\eps}^I(\s) &=&
{k^I\over 2\sqrt{\pi T}}
\lt(d(q, \eps)-d(q^{-1},\eps) +1 \rt) e^{i\vec k.\vec
  X}_{\eps}(\s') \cr
&& + 2 \lt[\Pi^I_{(-)}(\s-i\eps/2)e_{\eps}^{i\vec k.\vec X}(\s')
+ e_{\eps}^{i\vec k.\vec X}(\s') \Pi^I_{(+)}(\s+i\eps/2)\rt]~, \cr &&
\eea
where the notation $d(q,\eps)$ has been explained below
eqs.(\ref{reg-fields}). The term in the square bracket is the ordered part and will drop out when we anti-symmetrize and take $\eps \to 0$ limit. To find the contribution of the first term one uses the following identity\cite{kazama}\footnote{One way to derive this identity is to use (\ref{D-result}) and the result,
\bea
\delta_{\eps}(\Delta) \del_{\D}D_{\eps}(\Delta) = \lt( {1\over
  \pi}{\eps \over \Delta^2 + \eps^2}\rt)  \lt(-{1\over 2\pi T} {\Delta
  \over \Delta^2 + \eps^2} \rt) = {1\over 4\pi T}
\delta'_{\eps}(\Delta)~.
\label{delta-delD}
\eea}:
\bea
\delta_{\eps}(\Delta) \lt(d(q,\eps) -
d(q^{-1},\eps) \rt)
&=& -i\delta_{\eps}'(\Delta)~.
\eea
Using these results and the identity (\ref{identity}) in
(\ref{A-forms}) one finally gets the result for $A^R(\s,\s')$ in
(\ref{A-results}).

Let us now consider $A^L(\s,\s')$. MNO and PNO act differently in this sector. Going through the similar procedure as above one gets the following expressions in the two cases,
\bea
A^L(\s,\s') &\stackrel{\rm MNO}{\to}& \int {d\vec k\over (2\pi)^{d/2}}
\tilde K(\vec k) \lt[{\vec k^2  \over 2T} i\delta_{\eps}'(\Delta)
\lt(e_{\eps}^{i\vec k.\vec X}(\s')
\chi^2(\s') + e^{i\vec k.\vec X}_{\eps}(\s') \chi^2(\s) \rt) \rt. \cr
&& \lt. + 2\sqrt{\pi \over T} \delta_{\eps}(\Delta) k^I \lt\{\tilde
\Pi^I_{(-)}(\s+i\eps/2)e^{i\vec k.\vec X}_{\eps}(\s')\chi^2(s') +
e^{i\vec k.\vec X}_{\eps}(\s')\chi^2(s') \tilde
\Pi^I_{(+)}(\s-i\eps/2)\rt. \rt. \cr
&& \lt. \lt. -\tilde \Pi^I_{(-)}(\s'+i\eps/2)e^{i\vec k.\vec
  X}_{\eps}(\s)\chi^2(s) - e^{i\vec k.\vec X}_{\eps}(\s)\chi^2(s)
\tilde \Pi^I_{(+)}(\s'-i\eps/2) \rt\} \rt]~, \cr
&\stackrel{\rm PNO}{\to}& \int {d\vec k\over (2\pi)^{d/2}} \tilde
K(\vec k) \lt[-{\vec k^2
  \over 2T} i\delta_{\eps}'(\Delta) \lt(e_{\eps}^{i\vec k.\vec X}(\s')
\chi^2(\s') + e^{i\vec k.\vec X}_{\eps}(\s') \chi^2(\s) \rt) \rt. \cr
&& + 2\sqrt{\pi \over T} \delta_{\eps}(\Delta) k^I \lt\{\tilde
\Pi^I_{(+)}(\s-i\eps/2)e^{i\vec k.\vec X}_{\eps}(\s')\chi^2(s') +
e^{i\vec k.\vec X}_{\eps}(\s')\chi^2(s') \tilde
\Pi^I_{(-)}(\s+i\eps/2)\rt. \cr
&& \lt. \lt. -\tilde \Pi^I_{(+)}(\s'-i\eps/2)e^{i\vec k.\vec
  X}_{\eps}(\s)\chi^2(s) - e^{i\vec k.\vec X}_{\eps}(\s)\chi^2(s)
\tilde \Pi^I_{(-)}(\s'+i\eps/2) \rt\} \rt]~,
\label{AL-intermediate}
\eea
where we have already taken $\eps \to 0$ limit for some terms where it can be done unambiguously. Notice that the two expressions on the
right hand sides of  (\ref{AL-intermediate}) are equal at a finite
$\eps$. This can be proved as an identity. The only difference is that we have ordered infinite number of terms in two different ways. This is why the overall coefficients for the terms kept in the round brackets differ by a sign in the two cases. However, in
the $\eps \to 0$ limit $\s$ and $\s'$ coincide and the four terms kept in the curly brackets cancel each other in both the cases leading to two different results which differ by a sign. Therefore
depending on which expression we apply the $\eps \to 0$ limit to we
get the following results for the anomaly:
\bea
A^L(\s,\s') &\stackrel{\rm MNO}{\to}& -{i\over 4T} \lt(2 \vec \del^2
\dt \T_B(\s)
\delta'(\Delta) + \del_{\s} \vec \del^2 \dt \T_B(\s) \delta(\Delta)
\rt)~, \cr
&\stackrel{\rm PNO}{\to}& {i\over 4T} \lt(2 \vec \del^2 \dt \T_B(\s)
\delta'(\Delta) + \del_{\s} \vec \del^2 \dt \T_B(\s) \delta(\Delta)
\rt)~.
\label{AL-result}
\eea
Comparing the result for $A^R(\s,\s')$ (in (\ref{A-results})) and
(\ref{AL-result}) we see that for PNO $A^R(\s,\s')$ and $A^L(\s,\s')$ are same, but for MNO they differ by a sign. Therefore according to the first two equations in (\ref{anomalies}) conformal invariance would require us to have $[\dt \T_F(\s), \dt \T_F(\s')] = 0$ if
we adopt the MNO-prescription leading to a wrong space-time equation
of motion. We will see later in this section that similar ambiguity
exists in the fermionic sector and the MNO-prescription there will
indeed dictate this result.

Considering the third anomaly term in (\ref{A-results}) tells us that PNO-prescription is in fact the natural one to consider given our definition of the EM tensor. In this case the analogue of the terms in the curly brackets in (\ref{AL-intermediate}) do not cancel each other, rather it gives an ordered product of $k^I (\Pi^I-\tilde \Pi^I)$ and the exponential operator. Using eqs.(\ref{def-Pi}) such operators can be written as $\del_{\s}e^{i\vec k.\vec X}(\s)$. It is only when we consider PNO such a term cancels with the second term of the last equation in (\ref{A-forms}) which is already defined with PNO\footnote{Notice that
  irrespective of which normal ordering we consider in expressions
  like (\ref{AL-intermediate}), the
  exponential operators are always ordered according to PNO. This is
  because of the particular way the quantum EM tensor has been defined
  here.}. Because of this cancelation the final result turns out to
be the one given in eqs.(\ref{A-results}).

\vspace{.3in}
\noindent
{\bf \large Fermionic sector}

\noindent
Let us now turn to the fermionic sector. We will show that one
encounters a similar ambiguity in computing the commutator
in (\ref{expected}). To see how this ambiguity arises let us first
define,
\bea
F(\s) = S(\s)\Sigma \tilde S(\s)~,
\eea
such that, $\dt \T_F = {i\mu \over 2\sqrt{\pi T}} \chi F$. Since
$\chi$ behaves as a c-number for this computation, we
essentially need to calculate the commutator of $F(\s)$. The way it
was calculated in \cite{kazama} (which is analogous to the method
discussed in the bosonic sector above) is to first write:
\bea
[F(\s), F(\s')] &=& \Sigma_{ab} \Sigma_{cd} \lt[ \{\tilde S^b(\s),\tilde
 S^d(\s')\} S^c(\s') S^a(\s) - \{S^a(\s), S^c(\s')\}
\tilde S^b(\s) \tilde S^d(\s') \rt]~, \cr &&
\eea
then to use the anti-commutator in (\ref{comm}) to deduce,
\bea
[F(\s), F(\s')] &=& \delta(\s-\s') \lt[ S^a(\s') S^a(\s) - \tilde
S^a(\s) \tilde S^a(\s') \rt]~.
\eea
Notice that, just like in the bosonic case, the presence of the delta function forces the fermions to be coincident. Therefore each of the
terms in the square bracket drops out classically. But quantum
mechanically it is only a normal ordered product that is supposed to
vanish leaving a potential c-number contribution. One gets the following results for the two prescriptions at a finite $\eps$,
\bea
[F(\s), F(\s')]_{\eps} &\stackrel{\rm MNO}{\to}& -
\delta_{\eps}(\s-\s') \lt[ :S_{\eps}^a(\s)
S_{\eps}^a(\s'): + :\tilde S_{\eps}^a(\s) \tilde S_{\eps}^a(\s'):
\rt]~, \cr
&\stackrel{\rm PNO}{\to}& - \delta_{\eps}(\s-\s') \lt[ \ant
S_{\eps}^a(\s) S_{\eps}^a(\s')
\ant + \ant \tilde S_{\eps}^a(\s) \tilde S_{\eps}^a(\s')\ant \rt] +i d
\delta'_{\eps}(\s-\s')~. \cr &
\label{FF-results}
\eea
These results can easily be checked by using eqs.(\ref{SS-PNO-MNO})
and (\ref{delta-delD}).

Just like in the bosonic case the equality of the two right hand sides in (\ref{FF-results}) is an algebraic identity.
However, in the $\eps \to 0$ limit each of the terms kept in the
square brackets goes to zero as they are normal
ordered coincident fermions. This leads to two different results:
\bea
[F(\s), F(\s')] &\stackrel{\rm MNO}{\to}& 0 ~, \cr
&\stackrel{\rm PNO}{\to}&id \delta'(\D)~.
\label{FF-results-final}
\eea
Using the PNO-prescription and the identity (\ref{identity}) one
derives the expected result in (\ref{expected}).

As mentioned earlier, the fact that we get different results following the MNO and PNO-prescriptions in the above method of computation actually reflects an ambiguity and is not an artifact of different choices of vacuum. This can be seen very easily in the fermionic sector by doing the calculation in an alternative method\footnote{I
  thank R. Shankar for a useful discussion on this point.}. In this
method we first define the modes of $F(\s)$ as follows:
\bea
F_n \equiv \oint F(\s) e^{-in\s} =  \sum_{m \in Z} \lt(S_m\Sigma
\tilde S_{m-n} \rt)~.
\label{Fmodes}
\eea
Then a straightforward calculation shows:
\bea
[F_p, F_q] &=& \sum_{n\in Z} \lt(S^a_n S^a_{p+q-n} -\tilde
S^a_{-n-p}\tilde S^a_{n-q} \rt) ~, \cr
&=& \delta_{p+q,0} \sum_{n\in Z} \lt(S^a_n S^a_{-n} - \tilde
S^a_{-n-p}\tilde S^a_{n+p} \rt)~.
\label{FpFq-intermediate}
\eea
When $p+q\neq 0$, the first term in the first step can be evaluated
as: $\sum_{n\in Z}S^a_n S^a_{p+q-n} = {1\over 2} \sum_{n\in Z}
\{S^a_n, S^a_{p+q-n}\} = 0$. Similarly the second term also gives
zero. However, when $p+q=0$, each term gives ${d\over 2} \sum_{n\in Z} 1 \to \infty$ and therefore the result is ambiguous. Since
the result should be a c-number the best way to calculate it is to
compute the vacuum expectation value by directly using (\ref{Fmodes}). Considering the MNO-vacua $|\eta \ra$ as defined in appendix \ref{a:quant} it is straightforward to show that,
\bea
\la \eta' |[F_p,F_{-p}]|\eta \ra = -dp \la \eta'|\eta\ra~.
\eea
Using this one gets the result obtained in the PNO-prescription in
(\ref{FF-results-final}).

\section{Conclusion}
\label{s:conclusion}

In conclusion, we have suggested an explanation of how to reconcile
the observation (\ref{KYresult}) made by KY in \cite{kazama} with
the expectation that the right vacuum of the theory should be the
one associated to MNO. The key point is that the fields appearing in
the space-time effective theory are in general related to the
corresponding sigma-model couplings through some field
redefinition. For our restricted off-shell ansatz this field
redefinition can simply be taken to be identity if the interaction
term is ordered according to PNO. This is supported by showing that
the resulting EM tensor reproduces the correct physical spectrum for
the on-shell background. A consequence of such a definition of the EM tensor is that the Virasoro anomaly terms need to be ordered according to PNO before taking the $\eps \to 0$ limit. This fixes certain ambiguity in computing the Virasoro anomaly in the present method of computation and correctly reproduces the effective equation of motion. We also pointed out that this prescription is not in contradiction with the fact that the right vacuum is the one associated to MNO. It will be interesting to understand the relevance of the normal ordering prescriptions for the kind of analysis done here and in \cite{kazama} for more complicated backgrounds.

\begin{center}
{\bf Acknowledgement}
\end{center}

I thank Rajesh Gopakumar, R. Shankar and Nemani V. Suryanarayana for useful discussion and Nathan Berkovits and Yoichi Kazama for useful communication.

\appendix

\sectiono{Semi-light-cone quantization in R-R plane-wave}
\label{a:quant}

Here we will review the basic steps of the relevant semi-light-cone
quantization and then define the two normal ordering prescriptions of our interest. We will follow the same notations and conventions for
the space-time and world-sheet indices and gamma matrices as in
\cite{metsaev01}. Furthermore,  we take the following $SO(8)$
decomposition of the 16-dimensional gamma matrices:
$\gamma^0 = -\bar \gamma^0 = \Iop_{16}$, $\gamma^9=\bar \gamma^9 =
\pmatrix{\Iop_8 & 0 \cr 0 & -\Iop_8}$ and $\gamma^I = \bar \gamma^I =
\pmatrix{0&\s^I_{a\dot a} \cr \bar \s^I_{\dot a a}& 0}$, where
$I=1,\cdots ,d (=8)$ and $\s^I = (\bar \s^I)^T$ are the real $SO(8)$
gamma matrices.

The classical world-sheet lagrangian density ${\cal L}$ is given by,
\bea
2\pi \alpha' {\cal L} &=& -{1\over 2} \sqrt{-g} g^{ab} \lt( 2\del_a
X^+\del_b X^- - \mu^2 X_I^2 \del_a X^+ \del_b X^+
+ \del_a X^I \del_b X^I \rt) \cr
&& -i \sqrt{-g} g^{ab}\del_b X^+ \lt(\bar \theta \bar \gamma^- \del_a
\theta + \theta \bar \gamma^- \del_a \bar \theta
+ 2i\mu \del_a X^+ \bar \theta \bar \gamma^- \Pi \theta \rt) \cr
&& + i \eps^{ab} \del_aX^+ \lt(\theta \bar \gamma^-\del_b \theta +
\bar \theta \bar \gamma^-\del_b \bar \theta \rt)~,
\label{calL}
\eea
where the complex Weyl spinor $\theta$ is related to real
Majorana-Weyl spinors $\theta^A$ $(A=1, 2)$ satisfying the kappa gauge condition $\bar \gamma^+ \theta^A = 0$ in the same way as in
\cite{metsaev01}.

The theory has constraints and after going through Dirac's procedure \cite{kazama} the non-trivial equal time commutators (obtained from Dirac brackets) among the full set of basic operators are given by,
\bea
[X^{\mu}(\s), P_{\nu}(\s')] = i\delta^{\mu}_{\nu} \delta(\s-\s')~,
\quad
\{ S^A_a(\s), S^B_b(\s') \} = 2\pi \delta^{AB} \delta_{ab}
\delta(\s-\s')~,
\label{comm}
\eea
where $P_{\mu}(\s) = {\delta S \over \delta
  (\del_{\tau}X^{\mu}(\s))}$, $S$ being the world-sheet action and the
classical definition of $S^A(\s)$ is given by,
\bea
S^A(\s) = \sqrt{2\pi^A(\s)\over \sqrt{\alpha'}} \bar \gamma^-
\theta^A(\s)~,
\label{def-S}
\eea
where $\pi^1= \Pi^+~, ~\pi^2 =  \tilde \Pi^+$ and\footnote{Our
  definitions of $\Pi^{\mu}$ and $\tilde \Pi^{\mu}$ have been
  interchanged with respect to that in \cite{mukhopadhyay08}.},
\bea
\Pi^{\mu} = \sqrt{\pi \over T} \eta^{\mu \nu} P_{\nu} - \sqrt{\pi
  T}\del_{\s}X^{\mu}~, \quad
\tilde \Pi^{\mu} = \sqrt{\pi \over T} \eta^{\mu \nu} P_{\nu} +
\sqrt{\pi T}\del_{\s}X^{\mu}~,
\label{def-Pi}
\eea
where $T={1\over 2\pi \alpha'}$ is the string tension.

Next we define MNO and PNO. To do that we first mode expand various
fields in the following way (we have renamed: $S^1 \rightarrow S$,
$S^2 \rightarrow \tilde S$):
\bea
\Pi^{\mu}(\s) = \sum_n \Pi^{\mu}_n e^{in\s} ~, &\quad & \tilde
\Pi^{\mu}(\s) = \sum_n \tilde \Pi^{\mu}_n e^{-in\s} ~,  \cr
S^a(\s) =\sum_n S^a_n e^{in\s} ~, & \quad & \tilde S^a(\s) = \sum_n
\tilde S^a_n e^{-in\s}~,
\label{mode-exp}
\eea
such that the equal time commutators read: $[\Pi^{\mu}_m,\Pi^{\nu}_n]
= [\tilde \Pi^{\mu}_m, \tilde \Pi^{\nu}_n] = m \eta^{\mu \nu}
\delta_{m+n,0}$, $\{S^a_m,S^b_n\}=\{\tilde S^a_m, \tilde S^b_n\}=
\delta_{ab} \delta_{m+n}$. Then we define two sets of vacuum states
$|\eta \ra$ and $|\eta \ra'$ ($\eta = I, \dot a$ refer to the vector
and conjugate spinor representations of the fermion zero modes
respectively) such that they are annihilated by the following sets of operators:
\bea
\{\Pi^{\mu}_n~,\tilde \Pi^{\mu}_n \} |\eta \ra = 0~, \forall n\geq 0~,
&& \{S^a_n~, \tilde S^a_n \} |\eta \ra~,  \forall n>0~, \cr
\{\Pi^{\mu}_n~,\tilde \Pi^{\mu}_{-n} \} |\eta \ra' = 0~, \forall n\geq
0~, && \{S^a_n~, \tilde S^a_{-n} \} |\eta \ra'~,  \forall n>0~.
\label{def-eta-eta'}
\eea
MNO and PNO are defined to be the oscillator normal ordering with
respect to $|\eta\ra$ and $|\eta\ra'$ respectively.

\sectiono{Regularization procedure and propagators}
\label{a:reg}

Here we will discuss the regularization procedure that
we adopt for our explicit computations. Any local operator ${\cal
  O}(\Pi^{\mu}(\s), \tilde \Pi^{\mu}(\s), S^a(\s), \tilde S^a(\s))$
constructed out of the phase space variables is regularized by
replacing the arguments by their regularized versions:
${\cal O}_{\eps}(\Pi_{\eps}^{\mu}(\s), \tilde \Pi_{\eps}^{\mu}(\s),
S_{\eps}^a(\s), \tilde S_{\eps}^a(\s))$ where,
\bea
\Pi_{\eps}^{\mu}(\s) &=& \Pi^{\mu}_{(+)}(\s+i\eps/2) +
\Pi^{\mu}_{(-)}(\s-i\eps/2)~, \cr \tilde \Pi^{\mu}_{\eps}(\s) &=&
\tilde \Pi^{\mu}_{(+)}(\s-i\eps/2) + \tilde
\Pi^{\mu}_{(-)}(\s+i\eps/2)~, \cr
S^a_{\eps}(\s) &=& S^a_0+S^a_{(+)}(\s+i\eps/2)+S^a_{(-)}(\s-i\eps/2)~,
\cr
\tilde S^a_{\eps}(\s) &=& \tilde S^a_0+\tilde
S^a_{(+)}(\s-i\eps/2)+\tilde S^a_{(-)}(\s+i\eps/2)~,
\label{reg-fields}
\eea
where the subscripts $(\pm)$ refer to the annihilation and creation
parts according to MNO. For example \cite{kazama}, $\{S^a(\s),
S^b(\s')\} \to \{S^a_{\eps}(\s), S^b_{\eps}(\s')\}= \delta^{ab}
\lt(d(q,\eps) + d(q^{-1},\eps) -1 \rt)$
$=2\pi \delta^{ab} \delta_{\eps}(\D)~,$
where $\Delta=\s-\s'$, $q=e^{i\D}$ and $d(q,\eps) = \sum_{n\geq 0}q^n e^{-n\eps} = {1\over 1-qe^{-\eps}}$. Important identities involving $d(q,\eps)$ have been summarized in appendix A of \cite{kazama}.

To compute the equal time propagator for the bosonic fields we first
mode expand: $X^{\mu}(\s) = \sum_n X^{\mu}_n e^{in\s}$. Then using
$X^I_n ={i\over 2n \sqrt{\pi T}} \lt(\Pi^I_n - \tilde
\Pi^I_{-n} \rt)$, $\forall n\neq 0$ we regularize the bosonic fields
as,
\bea
X_{\eps}^{\mu}(\s) &=& X^{\mu}_0  -{1\over 2\sqrt{\pi T}}\int d
\s~\Pi_{\eps}^{\mu}(\s) + {1\over 2\sqrt{\pi T}}\int d \s ~\tilde
\Pi_{\eps}^{\mu}(\s)~,
\label{X-reg}
\eea
where we use indefinite integrals. Finally, a straightforward
computation gives the following expected results,
\bea
X_{\eps}^{\mu}(\s) X_{\eps}^{\nu}(\s') = \ant X_{\eps}^{\mu}(\s)
X_{\eps}^{\nu}(\s')\ant = :X_{\eps}^{\mu}(\s) X_{\eps}^{\nu}(\s'): +
\eta^{\mu \nu} D_{\eps}(\Delta)~,
\label{D-def}
\eea
where $:~:$ and $\ant ~\ant $ denote MNO and PNO respectively and the propagator $\eta^{\mu \nu} D_{\eps}(\Delta)$ is given by,
\bea
D_{\eps}(\Delta) &=& {i\over 4\pi T} \int d\Delta \lt(
d(q,\eps) - d(q^{-1},\eps)\rt)~, \cr
&=& -{1\over 4\pi T} \ln (\Delta^2+\eps^2)~.
\label{D-result}
\eea
To derive the second line we expand $d(q,\eps)$ in small
$\eps -i\Delta$ to first show $\del_{\Delta}D_{\eps}(\Delta) =
-{1\over 2\pi T} {\Delta \over \Delta^2 +\eps^2}$, then integrate this result. Notice that the first equality in (\ref{D-def}) indicates that given any classical function $f(x)$, the corresponding unordered operator $f(X(\s))$ is same as the one ordered according to PNO.

For the fermionic sector one finds,
\bea
S^a(\s) S^b(\s') &=& :S^a(\s) S^b(\s'): + \delta^{ab}
\lt(d(q,\eps) -1 \rt)~, \cr
\tilde S^a(\s) \tilde S^b(\s') &=& :\tilde S^a(\s) \tilde S^b(\s'): +
\delta^{ab} \lt(d(q^{-1},\eps) -1 \rt)~, \cr
&=& \ant \tilde S^a(\s) \tilde S^b(\s')\ant + \delta^{ab}
\lt(d(q,\eps) -1 \rt)~,
\eea
The analogue of (\ref{D-def}) in this case is given by,
\bea
\ant \tilde S^a(\s) \tilde S^b(\s')\ant = :\tilde S^a(\s) \tilde
S^b(\s'): +4\pi iT \delta^{ab} \del_{\Delta} D_{\eps}(\Delta)~.
\label{SS-PNO-MNO}
\eea

\sectiono{Normal ordering constant}
\label{a:constant}

Here we will compute the normal ordering constant as shown in
eqs.(\ref{calT0}) which is required to compute the physical spectrum
in subsection \ref{ss:spectrum}. The Virasoro modes are defined by the following mode expansions: $\T(\s)= \sum_n L_n e^{in\s}$, $\tilde
\T(\s)= \sum_n \tilde L_n e^{-in\s}$. As usual, only the Virasoro zero modes have the normal ordering ambiguity. It is clear from the
definition of EM tensor given in section \ref{s:EMtensor} that the
normal ordering constants do not receive any contribution from the
interaction part $\dt \T$. Therefore below we will simply restrict
ourselves to the flat background.

Let us consider the expression of $\T^{(0)}(\s)$ as given in
(\ref{calT0}) with an arbitrary normal ordering constant $a$ instead
of $1$. One can then directly compute the equal time commutator to get the following result:
\bea
[\T^{(0)}(\s), \T^{(0)}(\s')] &=& \pi i \lt[{3d/2+2-24 \xi \over 6}
\delta'''(\D) - \lt(4 \T^{(0)}(\s) -4 a -{1\over 3} \rt)
\delta'(\D) \rt. \cr
&& \lt.  - 2 \del_{\s}\T^{(0)}(\s) \delta (\D) \rt] ~.
\eea
In terms of the modes $L^{(0)}_n= \oint \T^{(0)}(\s) e^{-in\s}$ it
corresponds to:
\bea
[L^{(0)}_m, L^{(0)}_n] = (m-n) L^{(0)}_{m+n} + \lt[{3d/2+2-24 \xi
  \over 12} m^3 - \lt({1\over 6} +2a\rt) m\rt] \delta_{m+n}~,
\eea
Therefore choosing,
\bea
a=1~, \quad \xi = -{1\over 2}~,
\eea
one recovers the Virasoro algebra in the standard form:
$[L^{(0)}_m, L^{(0)}_n] = (m-n) L^{(0)}_{m+n} + {c\over 12} (m^3-m)
\delta_{m+n}$, with $c=26$. Similar results hold for the left moving
sector as well.

\sectiono{Proof of eqs.(\ref{A-forms})}
\label{a:proof}

Here we will prove the results in (\ref{A-forms}). Let us first
define the following operators,
\bea
&& l = :\Pi^+\Pi^-:+ \xi \del^2_{\s} \ln \Pi^+~, \quad
t={1\over 2} :\Pi^I\Pi^I:~, \quad s = -{i\over 2} (S\del_{\s}S)~, \cr
&&\tilde l = :\tilde \Pi^+\tilde \Pi^-:
+ \xi \del^2_{\s} \ln \tilde \Pi^+~, \quad
\tilde t={1\over 2} :\tilde \Pi^I\tilde \Pi^I:~,
\quad \tilde s = {i\over 2} (\tilde S\del_{\s}\tilde S)~,
\label{calT0-parts}
\eea
such that $\T^{(0)} =l+t+s+1$ and $\Tt^{(0)}=\tilde l+ \tilde t+\tilde
s+1$. Using the regularization procedure discussed in appendix
\ref{a:reg} and the commutation relations below eqs.(\ref{mode-exp})
one derives the following results:
\bea
[l(\s), \dt \T_B(\s')] &=& \pi i \Pi^+(\s) \tilde
\Pi^+(\s') K(\vec X(\s')) \delta'(\D)~,\cr
[l(\s), \dt \T_F(\s')] &=& {\mu \over 2}
\sqrt{\pi \over T} \Pi^+(\s) \sqrt{\tilde \Pi^+(\s') \over \Pi^+(\s')}
(S\Sigma \tilde S(\s')) \delta'(\D)~, \cr
[s(\s), \dt \T_F(\s')] &=& {\mu \over 2} \sqrt{\pi \over T}
\chi(\s') \lt[(S(\s)\Sigma \tilde S(\s')) \delta'(\D) -
(\del_{\s}S\Sigma \tilde S(\s)) \delta (\D) \rt]~.\cr &&
\label{r-comm}
\eea
\bea
[\tilde l(\s), \dt \T_B(\s')] &=& -\pi i
\Pi^+(\s') \tilde \Pi^+(\s) K(\vec X(\s')) \delta'(\D)~,\cr
[\tilde l(\s), \dt \T_F(\s')] &=& - {\mu \over 2}
\sqrt{\pi \over T} \tilde \Pi^+(\s) \sqrt{\Pi^+(\s') \over \tilde
  \Pi^+(\s')} (S\Sigma \tilde S(\s')) \delta'(\D)~, \cr
[\tilde s(\s), \dt \T_F(\s')] &=& - {\mu \over 2} \sqrt{\pi \over T}
\chi(\s') \lt[(S(\s')\Sigma \tilde S(\s)) \delta'(\D) -
(S\Sigma \del_{\s} \tilde S(\s)) \delta (\D) \rt]~,\cr &&
\label{l-comm}
\eea
where we have suppressed the cutoff $\eps$. Using the above results
and the identity:
\bea
{\cal O}(\s')\delta'(\D) = {\cal O}(\s) \delta'(\D) +
\del_{\s} {\cal O}(\s) \delta(\D)~,
\label{identity}
\eea
one gets,
\bea
[l(\s) + s(\s), \dt \T(\s')] -(\s
\leftrightarrow \s') &=& -4\pi i \dt \T(\s) \delta'(\D) - 2\pi i
\del_{\s} \dt \T(\s) \delta (\D)~, \cr
[\tilde l(\s) + \tilde s(\s), \dt \T(\s')] -(\s
\leftrightarrow \s') &=& 4\pi i \dt \T(\s) \delta'(\D) + 2\pi i
\del_{\s} \dt \T(\s) \delta (\D)~, \cr &&
\label{T-Ttilde-parallel-results}
\eea
and,
\bea
[l(\s) + s(\s), \dt \T(\s')] + [\dt
\T(\s), \tilde l (\s') + \tilde s(\s')] &=& \pi i \chi^2(\s)
\del_{\s}K(\vec X(\s)) \delta (\D)~. \cr &&
\label{T-Ttilde-paralle-results2}
\eea
Using (\ref{T-Ttilde-parallel-results}) in the first two equations in (\ref{bosonic-anomalies}) one gets the first two equations in
(\ref{A-forms}). Using (\ref{T-Ttilde-paralle-results2}) in the last equation of (\ref{bosonic-anomalies}) one gets the last equation in (\ref{A-forms}).

\sectiono{Problem with PNO-vacuum}
\label{a:PNO}

In subsection \ref{ss:spectrum} we computed the correct physical
spectrum by expanding the string field in the transverse Hilbert
space ${\cal H}_T$ which is a Fock space built over the MNO-vacua
$|p, \eta\ra$. In the context of bosonic string theory it was shown
in \cite{mukhopadhyay08} that the spectrum contains negative dimensions, as expected, if we consider the PNO-vacua $|p, \eta\ra'$ instead. Below we will briefly indicate that this is still true in the case of superstrings.

Let us try to see what changes we need to make in the analysis in
subsection \ref{ss:spectrum}. Recall that $\dt \T$ has already
been ordered according to PNO. The free parts in (\ref{calT0})
can be reordered according to PNO. But this does not change the final expression as the difference between MNO and PNO ordered expressions
is a c-number which cancels between the bosons and fermions. Therefore the operator expressions in (\ref{calS2perp}) remain the same and so does the problem of diagonalizing ${\cal S}_2^{\neq 0}$. Hence we arrive at the same expression as in (\ref{calS2diag}). The only difference is that now we expand the string field
$|\Psi_{\perp}\ra$ in terms of the basis states $|\{N\},p,\eta\ra'$
which can be obtained by replacing
$\tilde \Pi^I_{-n} \to \tilde \Pi^I_n$ and
$\tilde S^a_{-n} \to \tilde S^a_n$ in (\ref{basis1}). Moreover the
transformation (\ref{U-S}) is not a Bogoliubov transformation any more as it does not mix up the creation and annihilation operators (as defined in PNO). The same is true for the bosonic counterpart as well \cite{mukhopadhyay08}. Therefore $|p, \eta \ra'$ remains a vacuum with respect to the new oscillators, i.e.:
\bea
\{\alpha^I_n, ~\tilde \alpha^I_{-n}, ~U^a_n, ~\tilde U^a_{-n} \}
|p,\eta \ra' = 0~.
\eea
Writing the left moving part of ${\cal S}_2^{\neq 0}$ as $\sum_{n>0}\lt[\tilde \alpha^I_n \tilde \alpha^I_{-n} - w_n \tilde U^a_n \tilde U^a_{-n} \rt]$ it is now easy to see that both the
bosonic and fermionic parts have negative eigenvalues in the Hilbert space spanned by the states $|\{N\},p,\eta\ra'$.

\end{document}